\documentclass[twoside,11pt]{article}

\usepackage{jmlr2e}
\usepackage{textcomp}
\usepackage{threeparttable}
\usepackage{lipsum}

\usepackage[activate]{microtype}
\let\OLDthebibliography\thebibliography
\renewcommand\thebibliography[1]{
  \OLDthebibliography{#1}
  \setlength{\parskip}{0.3pt}
  \setlength{\itemsep}{2pt plus 0.6ex}
}
\usepackage{multirow}

\ShortHeadings{N-HANS}{Liu, Keren and Schuller}
\firstpageno{1}

\begin{document}

\title{N-HANS: Introducing the Augsburg\\Neuro-Holistic Audio-eNhancement System}

\author{\name Shuo Liu \email shuo.liu@informatik.uni-augsburg.de \\
       \name Gil Keren \email gil.keren@informatik.uni-augsburg.de \\
       \name Bj\"orn Schuller \email schuller@ieee.org \\ 
       \addr ZD.B Chair of Embedded Intelligence for Health Care and Wellbeing, 
       University of Augsburg, Germany \\
       GLAM -- Group on Language, Audio, \& Music,
       Imperial College London, London, UK \\
       }
\editor{Editor's Name}

\maketitle

\vspace{-0.3cm}
\begin{abstract}
N-HANS is a Python toolkit for in-the-wild audio enhancement, including speech, music, and general audio denoising, separation, and selective noise or source suppression. The functionalities are realised based on two neural network models sharing the same architecture, but trained separately. The models are comprised of stacks of residual blocks, each conditioned on additional speech or environmental noise recordings for adapting to different unseen speakers or environments in real life.
In addition to a Python API, a command line interface is provided to researchers and developers, both of which are documented at \url{https://github.com/N-HANS/N-HANS}. Experimental results indicate that N-HANS achieves outstanding performance, and ensure its reliable usage in real-life audio and speech-related tasks, reaching very high audio and speech quality.  
\end{abstract}


\vspace{-0.3cm}
\section{Introduction}
\vspace{-0.25cm}
In real life, audio signals are prone to corrupt with background noise and interference, such as transportation noise, industrial noise, and voices of unwanted speakers, etc. Contaminated audio signals are not only disruptive to hearing perception, especially for the hearing impaired, but may severely weaken the performance of general audio and specific speech-driven applications, such as automatic speech recognition, or speech emotion recognition \citep{triantafyllopoulos2019towards,monaghan2017auditory,Keren2018}. Hence, audio enhancement, which generally aims at extracting desired audio signals, is broadly exploited to improve audio and speech quality for real-life applications.  
Audio denoising and separation are two specific tasks for audio enhancement, where audio denoising suppresses background noise from noisy audio signals, while audio separation attempts to extract target audio from a mixture of multiple overlapping audio signals that belong to different sources. Neural network based models for audio denoising \citep{liu2014experiments} and separation \citep{wang2018supervised,huang2014deep} have been reported in the literature to be effective and surpass some classic algorithms which are built on some additional assumptions \citep{Delic2019SpeechTP}.

The removal of noise components and interference from in-the-wild audio remains challenging in the following aspects. 
First of all, an audio enhancement model relies on its generalisability for noise and hence, is limited to the data size and diversity of the training noise. However, audio may simultaneously be corrupted by multiple kinds of noise, including some unseen noise. 
Secondly, removing noise or interference while preserving an audio signal of interest such as speech requires accurate estimation of the noise and interference components in the audio signal. However, aggressive estimation commonly happens for these tasks and therefore affects the audio naturalness such as speech's naturalness. 
Thirdly, some noisy audio samples may contain useful other signals like warnings, e\,.g., an aerial defence alarm. Extracting only certain target audio but ignoring other useful signals, in this case, could result in an unsafe setting. Moreover, in certain circumstances, only a specific noise could be desired to be removed from the audio, the other environmental and ambient noise can be preserved to pertain a natural audio surrounding. A selective noise suppression system for this purpose should allow to select desired noise, named as ``positive noise'' in the ongoing, and suppress undesired noise, named as ``negative noise''.
In addition, when interfering audio sources appear in noise, recovering from the same type of context becomes more challenging, especially if the target and interference share similar acoustic properties such as similar pronunciation in the case of speech. 

To best alleviate the above issues, we introduce our Tensorflow based Python toolkit Neuro-Holistic Audio-eNhancement System (N-HANS) for in-the-wild audio denoising, suppression, and selective audio separation. N-HANS is embedded with two trained models, which share an identical architecture, but separate parameters. For a noisy audio of interest, contaminated with environmental noise or interfering audio sources such as  speakers, additional recordings are demanded as input reference for noise adaptation or audio source adaptation. A positive recording implicates the contents to be preserved in the noisy audio, and a negative recording hints the contents to be suppressed. N-HANS can perform as front end to interface with openXBOW \citep{max2017} and auDeep \citep{shahin2018}, both of which have been broadly applied for audio features extraction.


\vspace{-0.35cm}
\section{\textpm Auxiliary Network \& System Overview}
\begin{figure*}[t!]
  \vspace{-1cm}
  \centering
  \includegraphics[width=22.5cm]{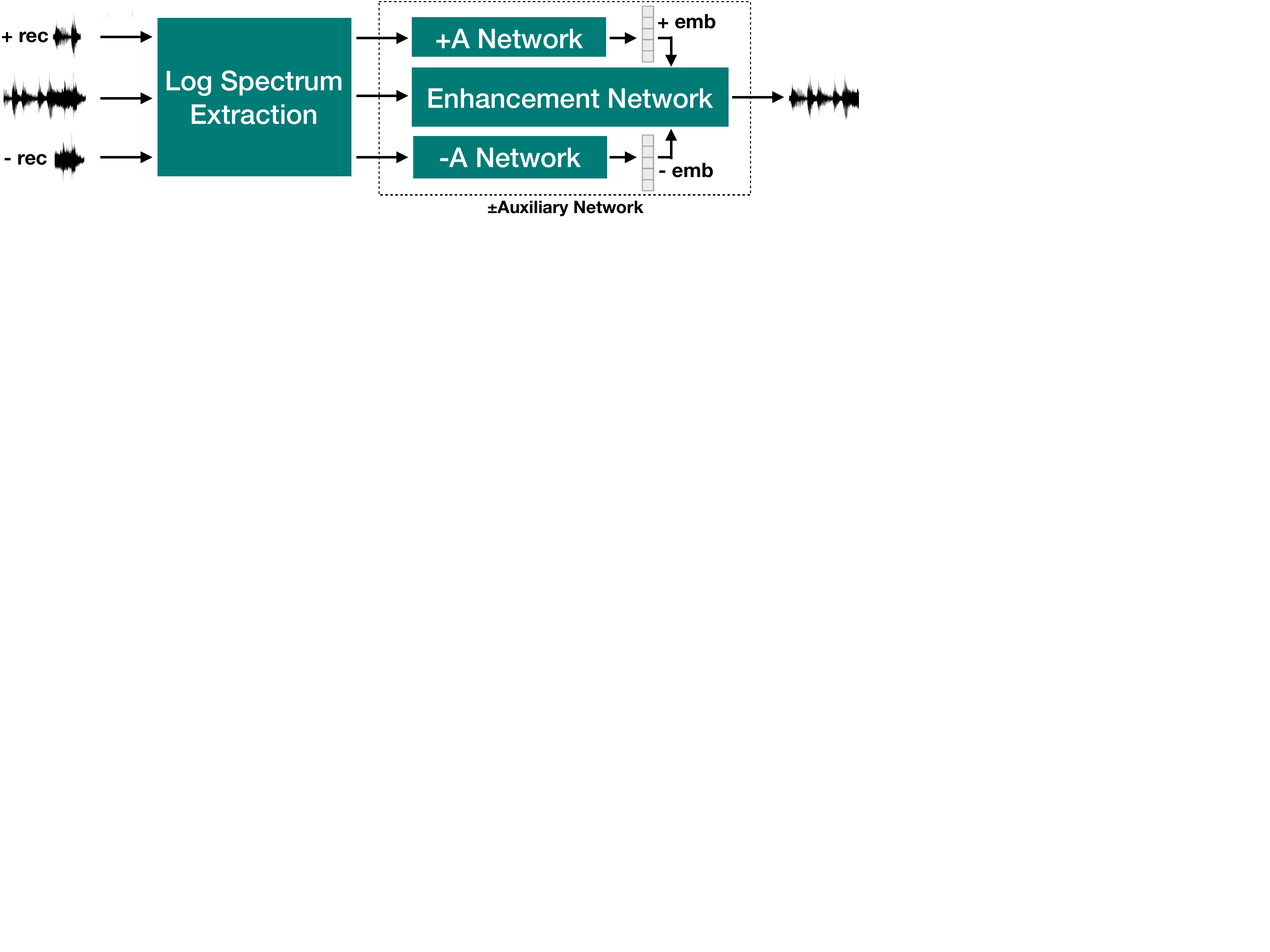}
  \vspace{-14cm}
  \caption{System framework and the \textpm Auxiliary Network. The $+$A Network processes the positive recording ($+$rec) to produce a positive embedding vector ($+$emb) that implicates the components, in noisy audio, to be preserved. The $-$A Network processes the negative recording ($-$rec) to obtain the negative embedding vector ($-$emb) that hints the components to be suppressed. The enhancement network processes the noisy audio, and the positive and negative embedding, to emit the desired audio.} 
  \vspace{-0.5cm}
  \label{fig:model}
\end{figure*}

\vspace{-0.25cm}
Since a selective noise suppression system attempts to remove negative noise, while preserving audio of interest and positive noise, an audio denoising system, which aims at achieving clean audio of interest is seen as a special case, when positive noise is set to mute. An audio source separation system separates a mixture audio to recover a target audio source, while removing interference audio sources. 
Since all three systems have a common goal that they suppress an undesired signal part, while preserving a desired signal part, we introduce the \textpm Auxiliary Network (\textpm A Net) as illustrated in Figure \ref{fig:model}.

\begin{table}[t!]
  \vspace{-1cm}
  \small
  \caption{N-HANS Input \& Output Information}
  \label{tab:inout}
  \centering
  \setlength{\tabcolsep}{1.7mm}
  \renewcommand{\arraystretch}{1}
  \begin{tabular}{ l  l  l  l l}
    \hline
    \multicolumn{1}{c}{\textbf{Task}} & 
                                         \multicolumn{1}{c}{\textbf{input}} & \multicolumn{1}{c}{\textbf{$+$recording}} & \multicolumn{1}{c}{\textbf{$-$recording}} &
                                         \multicolumn{1}{c}{\textbf{output}}\\
    \hline
    Selective Denoiser & noisy audio & noise to preserve & noise to suppress & denoised audio\\
    Denoiser & noisy audio & \hspace{1cm}- & noise to suppress & denoised audio\\
    Separator & overlapping source & target source & interference source & separated source\\
    \hline
  \end{tabular}
  \vspace{-0.4cm}
\end{table} 

First of all, the log magnitude spectrum of the input noisy audio, positive, and negative recordings are extracted by taking the log absolute value of the Short-Time Fourier Transformation (STFT), and fed to the \textpm A Network separately. Then, the $+$A Network processes the extracted positive spectrum to produce a positive embedding vector, while the $-$A Network processes the negative spectrum to emit a positive embedding vector. The positive and negative embedding vectors are then injected into the enhancement network to emit the denoised or separated audio. With awareness of positive and negative contexts, N-HANS achieves applicability to unseen noisy environments and audio sources such as speakers. The input and output information of the N-HANS system for different tasks refers to Table \ref{tab:inout}.


\vspace{-0.4cm}
\section{Practical Usage}
\vspace{-0.25cm}
N-HANS is implemented using Python2 and Tensorflow 1.14, and can be used on any common platform. Its source code has been published in GitHub as a public repository. In addition, for users who want to directly apply N-HANS, trained models are included in the repository, which can also be downloaded via a command line interface. 
N-HANS supports the Waveform Audio File Format (WAV) for input and output, which is one of the most standard and broadly used audio formats.  

With the command line interface, N-HANS can be applied for a single audio example, or for multiple audio examples organised into one directory. Although N-HANS is suggested to run with GPU-acceleration, it is capable of running with CPU only. It performs audio denoising, selective denoising or separation faster than real-time, taking about $0.272$ seconds to operate on one second of audio using a single NVIDIA Titan X Pascal GPU, hence roughly resembling a Real-Time-Factor (RTF) of 3.7.  

\begin{table}[t!]
  \vspace{-0.25cm}
  \small
  \caption{LibriSpeech \& AudioSet Test Evaluation Metrics for N-HANS Denoising.}
  \label{tab:denoiser}
  \centering
  \setlength{\tabcolsep}{4.5mm}
  \renewcommand{\arraystretch}{1}
  \begin{tabular}{ l r r r r r r}
    \hline
    \multicolumn{1}{c}{\textbf{SNR}} &  \multicolumn{1}{c}{\textbf{LSD}} &                                          \multicolumn{1}{c}{\textbf{SDR}} &
                                        \multicolumn{1}{c}{\textbf{PESQ}} &
                                        \multicolumn{1}{c}{\textbf{STOI}} &
                                        \multicolumn{1}{c}{\textbf{MCD}} &
                                        \multicolumn{1}{c}{\textbf{SSNR}} \\
    \hline
    $0\,dB$ & $1.17$ & $7.02$ & $1.58$ & $0.81$ & $6.79$ & $4.06$\\
    $3\,dB$ & $1.10$ & $8.72$ & $1.74$ & $0.84$ & $6.51$ & $5.10$\\
    $5\,dB$ & $1.05$ & $9.60$ & $1.62$ & $0.86$ & $6.40$ & $5.90$\\
    $10\,dB$ & $0.93$ & $11.86$ & $1.80$ & $0.90$ & $5.98$ & $7.80$\\
    $15\,dB$ & $0.84$ & $13.35$ & $1.92$ & $0.92$ & $5.49$ & $9.58$\\
    \hline
  \end{tabular}
  \vspace{-0.4cm}
\end{table}

\vspace{-0.4cm}
\section{Experiments}
\vspace{-0.25cm}
We conducted experiments to test N-HANS as a speech denoising system and selective noise suppression system with the LibriSpeech \citep{librispeech} and AudioSet \citep{Gemmeke2017AudioSet} databases, which provide large-scale clean speech and real-life noise respectively. LibriSpeech corpus contains its own training, development, and test splits, and the split of AudioSet corpus is given in the Github repository, according to all categories in AudioSet's ontology. Samples from Librespeech and AudioSet are mixed to create noisy speech samples. 
For selective noise suppression, two samples from AudioSet are selected as positive and negative recordings for each noisy sample. 
Test set evaluation metrics, log spectral distortion (LSD), signal-to-distortion ratio (SDR), perceptual evaluation of speech quality (PESQ), short-time objective intelligibility (STOI), mel cepstral distortion (MCD), and segmental SNR (SSNR), which are widely used in prior works such as \citep{jeon2017audio}, are demonstrated in Table \ref{tab:denoiser} and Table \ref{tab:selective} for different input signal-to-noise ratio (SNR) cases.
 
\begin{table}[t!]
  \vspace{-1.2cm}
  \caption{LibriSpeech \& AudioSet Test Evaluation Metrics for N-HANS Selective Denoising. $+$SNR represents SNR for positive noise, and $-$SNR for negative noise.}
  \label{tab:selective}
  \small
  \centering
  \setlength{\tabcolsep}{3mm}
  \renewcommand{\arraystretch}{1}
  \vspace{-0.8\baselineskip}
  \begin{tabular}{ l r r r r r r r}
    \hline
    \multicolumn{1}{c}{\textbf{$+$SNR}} & \multicolumn{1}{r}{\textbf{$-$SNR}} &
                                          \multicolumn{1}{r}{\textbf{LSD}} & 
                                          \multicolumn{1}{r}{\textbf{SDR}} & 
                                          \multicolumn{1}{r}{\textbf{PESQ}} &
                                          \multicolumn{1}{r}{\textbf{STOI}} &
                                          \multicolumn{1}{r}{\textbf{MCD}} &
                                          \multicolumn{1}{r}{\textbf{SSNR}} \\
    \hline
    \multirow{4}{*}{\textbf{$0\,dB$}} & \textbf{$0\,dB$} & $0.76$ & $7.72$ & $1.88$ & $0.79$ & $5.36$ & $7.13$ \\
                                     & \textbf{$3\,dB$} & $0.69$ & $9.49$ & $1.93$ & $0.84$ & $4.96$ & $8.55$ \\
                                     & \textbf{$5\,dB$} & $0.65$ & $10.64$ & $1.97$ & $0.87$ & $4.67$ & $9.46$ \\
                                     & \textbf{$8\,dB$} & $0.59$ & $12.12$ & $1.94$ & $0.90$ & $4.29$ & $10.83$ \\
    \hline
    \multirow{4}{*}{\textbf{$3\,dB$}} & \textbf{$0\,dB$} & $0.78$ & $7.16$ & $1.89$ & $0.78$ & $5.58$ & $6.56$\\
                                     & \textbf{$3\,dB$} & $0.73$ & $8.93$ & $1.91$ & $0.83$ & $5.29$ & $7.79$\\
                                     & \textbf{$5\,dB$} & $0.68$ & $10.06$ & $1.98$ & $0.85$ & $4.97$ & $8.90$\\
                                     & \textbf{$8\,dB$} & $0.64$ & $11.46$ & $1.94$ & $0.88$ & $4.68$ & $9.93$\\
    \hline
    \multirow{4}{*}{\textbf{$5\,dB$}} & \textbf{$0\,dB$} & $0.81$ & $7.19$ & $1.82$ & $0.78$ & $5.71$ & $6.30$\\
                                     & \textbf{$3\,dB$} & $0.75$ & $8.68$ & $1.88$ & $0.82$ & $5.44$ & $7.44$\\
                                     & \textbf{$5\,dB$} & $0.72$ & $9.76$ & $1.85$ & $0.84$ & $5.28$ & $8.27$\\
                                     & \textbf{$8\,dB$} & $0.65$ & $11.33$ & $1.90$ & $0.88$ & $4.83$ & $9.85$\\
    \hline
    \multirow{4}{*}{\textbf{$8\,dB$}} & \textbf{$0\,dB$} & $0.86$ & $7.01$ & $1.78$ & $0.77$ & $5.99$ & $5.83$\\
                                     & \textbf{$3\,dB$} & $0.79$ & $8.63$ & $1.82$ & $0.81$ & $5.71$ & $7.07$\\
                                     & \textbf{$5\,dB$} & $0.74$ & $9.62$ & $1.88$ & $0.84$ & $5.44$ & $7.92$\\
                                     & \textbf{$8\,dB$} & $0.68$ & $11.30$ & $1.89$ & $0.87$ & $5.09$ & $9.36$\\
    \hline
  \end{tabular}
     
  \vspace{-0.2cm}
\end{table}

The performance of N-HANS as a speech separation system and two state-of-the-art baselines \citep{hershey2016deep, huang2014deep} recently proposed for source separation are compared using the large and diverse VoxCeleb data set \citep{DBLP:conf/interspeech/NagraniCZ17,DBLP:conf/interspeech/ChungNZ18}, which contains two versions, each contains its own training and test test. Three objective evaluation metrics, signal-to-distortion ratio (SDR), signal-to-artifacts ratio (SAR), and signal-to-interference ratio (SIR) \citep{vincent2006performance} are compared in Table \ref{tab:separator}.

\begin{table*}[t!]
  \vspace{-0.2cm}
  \caption{VoxCeleb Test Set Evaluation Metrics for N-HANS Speech Separation. ``f" stands for female, and ``m" for male, indicating the mixture cases of different genders.}
  \label{tab:separator}
  \centering
  \setlength{\tabcolsep}{0.45mm}
  \renewcommand{\arraystretch}{1}
  \vspace{-0.8\baselineskip}
  \begin{tabular}{ l r r r r r r r r r r r r r r r}
  \hline
    \multicolumn{1}{c}{\textbf{}} &
    \multicolumn{1}{c}{\textbf{}} &
    \multicolumn{4}{c}{\textbf{SDR}} &
    \multicolumn{1}{c}{\textbf{}} &
    \multicolumn{4}{c}{\textbf{SAR}} & 
    \multicolumn{1}{c}{\textbf{}} &
    \multicolumn{4}{c}{\textbf{SIR}}\\
    \multicolumn{1}{c}{\textbf{Method}} &
    \multicolumn{1}{c}{\textbf{}}&
    \multicolumn{1}{c}{\textbf{f+f}} & \multicolumn{1}{c}{\textbf{m+m}} & \multicolumn{1}{c}{\textbf{f+m}} & 
    \multicolumn{1}{c}{\textbf{all}} &
    \multicolumn{1}{c}{\textbf{}} &
    \multicolumn{1}{c}{\textbf{f+f}} & \multicolumn{1}{c}{\textbf{m+m}} &
    \multicolumn{1}{c}{\textbf{f+m}} &
    \multicolumn{1}{c}{\textbf{all}} &
    \multicolumn{1}{c}{\textbf{}} &
    \multicolumn{1}{c}{\textbf{f+f}} & \multicolumn{1}{c}{\textbf{m+m}} & \multicolumn{1}{c}{\textbf{f+m}} &
    \multicolumn{1}{c}{\textbf{all}} \\
    \hline
    \textbf{Deep Clustering}    
    &  & 0.51 & 0.64 & 1.12 & 0.84 & & 1.39 & 2.01 & 2.35 & 2.09 & & 5.66 & 5.64 & 7.73 & 6.58 \\
    \textbf{Deep Attractor}   
    & & 1.43 & 1.17 & 2.53 &1.81 & & 2.64 & 2.93 & 3.50 & 3.29 & & \textbf{8.27} & \textbf{9.12} & \textbf{12.22} & \textbf{10.41} \\
    \textbf{N-HANS} & & \textbf{4.76} & \textbf{4.55} & \textbf{5.03} & \textbf{4.79} & & \textbf{8.70} & \textbf{8.05} & \textbf{8.75} & \textbf{8.44} & & 6.69 & 6.82 & 7.50 & 7.11 \\
  \hline
  \end{tabular}
  \vspace{-0.3cm}
\end{table*}

\vspace{-0.35cm}
\section{Conclusions and Outlook}
\vspace{-0.25cm}
N-HANS is an open source toolkit developed based on our proposed \textpm Auxiliary Network for audio denoising, source separation, and selective noise suppression. Conditioning on reference recordings, N-HANS is capable of adapting to unseen environments and audio sources such as speakers. Future work for audio enhancement should focus on improving speech intelligibility in extreme low SNR cases, to overcome the distortions that occasionally introduced in audio. For general audio separation, more work will be done to extend the system to any number of audio sources.

This project has received funding from the European Union’s Horizon 2020 research and Innovation programme under grant agreement No. 688835 (RIA DE-ENIGMA)


\newpage

\vskip 0.2in
\bibliographystyle{IEEEtran}
\bibliography{sample}

\end{document}